\documentclass[pre,twocolumn]{revtex4}

\usepackage{amsmath}

\newcommand{\avg}[1]{\ensuremath{\langle #1 \rangle}}
\newcommand{\D}{\ensuremath{\mathrm{d}}}
\newcommand{\rme}{\ensuremath{\mathrm{e}}}
\newcommand{\jroot}{\ensuremath{\alpha_\nu^n}}

\begin{document}

\title{Survival of a diffusing particle in an expanding cage}
\date{\today}
\author{Alan J. Bray}
\author{Richard Smith}
\affiliation{School of Physics and Astronomy, The University of Manchester, 
Manchester M13 9PL, U.K.}

\begin{abstract}
We consider  a Brownian particle, with diffusion  constant $D$, moving
inside  an  expanding  $d$-dimensional  sphere  whose  surface  is  an
absorbing  boundary for the  particle. The  sphere has  initial radius
$L_0$  and expands  at a  constant  rate $c$. We calculate the joint
probability density, $p(r,t|r_0)$, that the  particle survives until 
time $t$, and is at a distance $r$ from  the centre of the sphere, 
given that it started at  a distance $r_0$ from  the centre. The  
asymptotic ($t \to \infty$) probability, $Q$, obtained  by integrating  
over   all  final positions, that the particle survives, starting from 
the centre of the sphere,  is given by  
$Q = [4/\Gamma(\nu+1)\lambda^{\nu+1}] \sum_n
b_n\,\exp[-(\alpha_\nu^n)^2/\lambda]$, where  $\lambda = cL_0/D$, $b_n
=  (\alpha_\nu^n)^{2\nu}/[J_{\nu+1}(\alpha_\nu^n)]^2$,  $\nu=(d-2)/2$,
and $\alpha_\nu^n$ is  the $n$th positive zero of  the Bessel function
$J_\nu(z)$. The cases  $d=1$ and $d=3$ are especially  simple, and may
be solved elegantly using backward Fokker-Planck methods.

\end{abstract}

\maketitle

\section{Introduction}
First-passage problems  for stochastic systems  have attracted renewed
interest  in  recent  years  \cite{redner}, notably  in  systems  with
infinitely many coupled degrees of freedom, where the effective
stochastic  process describing  a single  degree of  freedom  is often
non-Markovian  \cite{majumdar}.  However,  even  in systems  with  few
degrees  of  freedom, and  Markovian  dynamics,  problems with  moving
boundary conditions have proved difficult to solve.

In this  paper we consider a single diffusing particle, or Brownian walker,
moving within a $d$-dimensional sphere, of initial radius $L_0$, which
is expanding at a constant rate $c$. The surface of the sphere defines
an absorbing boundary  for the particle. We present exact results, for  
any space dimension $d$, for the probability that the particle survives  
up to  time $t$. 

In an  earlier paper  \cite{1dpaper} this problem  was solved  for the
special  case $d=1$, in  the limit  $t \to  \infty$, using  an elegant
method  based  on  the  backward Fokker-Planck  equation.  In  section
\ref{sec:BFPE} we recap  how this method works and show  how it can be
extended to the  case $d=3$. The method  does  not, however, appear to
be useful for other values of $d$. Why this should be so becomes clear
later in the paper, where we obtain a complete solution for general $d$.

To   our    knowledge,   the   results   presented    here   (and   in
\cite{1dpaper}) are the  first exact results for this type of problem.
Approximate  methods  have  been  developed by  Krapivsky  and  Redner
\cite{krap}  in the  limit of  slow (``adiabatic  approximation'') and
fast (``fast  approximation'') motion  of the absorbing  boundary. The
relevant dimensionless  parameter is $\lambda = cL_0/D$,  where $D$ is
the  diffusion constant  of the  particle.  The slow  and fast  limits
correspond to small and large $\lambda$ respectively.

In section \ref{sec:FFPE} we  show that the adiabatic approximation of
Krapivsky and  Redner can be modified  to obtain an  exact solution of
the   usual  (forward)   Fokker-Planck  equation   for   the  survival
probability of the particle for any time $t$, and any initial position
within the sphere, for any space dimension $d$.

The outline of the paper  is as follows. In section \ref{sec:BFPE}, we
present  the  backward Fokker-Planck  equation  for the  infinite-time
survival  probability, and  its  solutions for  $d=1$  and $d=3$.   In
section \ref{sec:FFPE} we show  how the forward Fokker-Planck equation
can  be  solved  to  obtain  the survival  probability  in  any  space
dimension, for any time $t$ and for an arbitrary starting point within
the sphere. In section \ref{sec:infinite} we extract the infinite-time
survival  probability. The previous  results for  $d=1$ and  $d=3$ are
recovered as special cases. For general $d$, it is not straightforward
to  extract  analytically,  from  the  exact  solution,  the  limiting
behaviour for large $\lambda$,  so in section \ref{sec:fast} we employ
the  ``fast  approximation'' of  Krapivsky  and  Redner  to obtain  an
approximate  solution  that  becomes  exact  in  this  limit.  Section
\ref{sec:summary} concludes with a brief summary of the results.

\section{Infinite time survival probability of a particle in a 
expanding   cage: The backward Fokker-Planck method}  
\label{sec:BFPE}

\subsection{Solution in One Dimension}

In  \cite{1dpaper}  we considered  a  diffusing  particle obeying  the
Langevin equation  $\dot{X}(t)=\eta(t)$, $X(0)=x$, where  $\eta(t)$ is
Gaussian   white   noise   with   mean  zero   and   time   correlator
$\avg{\eta(t_1)  \eta(t_2)}  =   2D  \delta(t_1-t_2)$,  bounded  by  a
linearly--expanding absorbing  cage with edges located at  $\pm (L_0 +
ct)$. After a  time $\Delta t$ the boundaries  have moved to positions
$\pm (L_0 + c(t+\Delta t)$, and  the particle has moved to $x + \Delta
x$, where  $\langle \Delta x \rangle  = 0$, and  $\avg{(\Delta x)^2} =
2D\,\Delta t$.  The probability, $Q(x,L_0,t)$ that  the particle still
survives after a time $t$  satisfies the obvious equation $Q(x,L_0,t) =
\avg{Q(x+\Delta x,  L_0+c\Delta t,  t-\Delta t)}$. Expanding  to first
order in $\Delta t$ yields the backward Fokker-Planck equation
\begin{equation}
\frac{\partial Q}{\partial t} = D \frac{\partial^2 Q}{\partial x^2}
+ c \frac{\partial Q}{\partial L_0}. 
\end{equation}  
The novelty  in this approach  resides in treating $L_0$,  which gives
the initial positions of  the boundaries, as an additional independent
variable.

Discarding   the  time-derivative   term,  to   directly   obtain  the
infinite-time   limit,   and   introducing   dimensionless   variables
$y=cx/D$ and $\lambda = cL_0/D$ gives the simplified equation
\begin{equation}
\label{1bfpe}
\frac{\partial^2 Q}{\partial y^2} + \frac{\partial Q}{\partial
\lambda}=0,
\end{equation}
subject    to    the     absorbing    boundary    conditions    $Q(\pm
\lambda,\lambda)=0$, and with $Q(y,\lambda \to \infty)=1$. The solution 
is
\begin{equation}
\label{coshsoln}
Q(y,\lambda)=  \sum_{n=-\infty}^\infty  (-1)^n  \cosh (ny)  \rme^{-n^2
\lambda}, 
\end{equation}
which satisfies both the differential equation and the boundary 
conditions. 

For  a particle  starting at the origin ($y=0$), the survival
probability $Q(0,\lambda)$ is given by
\begin{equation}
\label{qlambda}
Q(0,\lambda)= \sum_{n=-\infty}^\infty (-1)^n \rme^{-n^2 \lambda} \sim
1-2\rme^{-\lambda }, \; \; \lambda \to \infty.
\end{equation}

We can extract the small-$\lambda$ behaviour by 
 using the Poisson sum formula,
\begin{displaymath}
\sum_{n=-\infty}^\infty   f(n)=\sum_{k=-\infty}^\infty  \tilde{f}(2\pi
k),
\end{displaymath}
where  $\tilde{f}(k)$ is  the Fourier  transform of  the  the function
$f(n)$, to recast Eq.~(\ref{coshsoln}) in the form
\begin{equation}
\label{poisson}
Q(0,\lambda)=\sqrt{\frac{\pi}{\lambda}} \sum_{k=-\infty}^\infty
\rme^{-\pi^2 (2k-1)^2 /4 \lambda} \sim
2\sqrt{\frac{\pi}{\lambda}}\rme^{-\frac{\pi^2}{4\lambda}}, \; \lambda
\to 0.
\end{equation}

\subsection{Solution in Three Dimensions}

We now  extend this  result  to the  case  of a  diffusing
particle    in    three    spatial    dimensions   bounded    by    an
linearly--expanding,  absorbing sphere  of radius  $L(t)=L_0+ct$ using
the  same  backward  Fokker--Planck  method.  


For general spatial dimensionality $d$, equation (\ref{1bfpe}) has the 
obvious generalisation
\begin{equation}
\nabla^2 Q + \frac{\partial Q}{\partial \lambda} = 0
\end{equation}
where  the  dimensionless  spatial  coordinate  is ${\bf  r}  =  c{\bf
r}_0/D$, and  $\lambda = cL_0/D$ as  before, where ${\bf  r}_0$ is the
initial  location of the  particle within  the sphere.  Exploiting the
spatial isotropy, we infer that  $Q$ depends on ${\bf r}$ only through
its magnitude, $r = |{\bf r}|$, giving, for $d=3$,
\begin{equation}
\label{3bfpe}
\frac{\partial^2 Q}{\partial r^2} +\frac{2}{r} 
\frac{\partial Q}{\partial r} + \frac{\partial Q}{\partial \lambda}=0,
\end{equation}
This equation has separable solutions of the form 
\begin{equation*}
Q_k(r,\lambda )=\frac{\sinh (kr)}{r}\; \rme^{-k^2 \lambda}, 
\end{equation*}
characterised by  an index $k$, from  which a general  solution can be
constructed  by  superposition.  Note   that  solutions  of  the  form
$[\cosh(kr)/r]\exp(-k^2\lambda)$ are rejected  as they are not regular
at the origin.  At this point we make no  assumptions about the values
of $k$.  We find, however,  that no simple superposition  of solutions
$Q_k(r,\lambda)$    satisfies   the   required    boundary   condition
$Q(\lambda,\lambda  )=0$. We  note,  however, that  any derivative  of
$Q_k(r,\lambda)$  with   respect  to  $k$   is  also  a   solution  of
Eq.~(\ref{3bfpe}) since it is simply  a superposition of two values of
$k$ which are infinitesimally  close together. We therefore try taking
the  first derivative  of this  solution and  postulate a  sum  of the
functions $dQ_k(r,\lambda)/dk$ with integer  values of $k$, in analogy
with the one-dimensional case, to obtain
\begin{equation*}
Q(r,\lambda )= \sum_{n=-\infty}^\infty a_n \left[ \cosh (nr)
-\frac{2n\lambda}{r}\sinh (nr) \right] \rme^{-n^2 \lambda }.
\end{equation*}
If  we choose the  amplitudes to  be $a_k=1$  for all  $k$, we  see by
inspection  that the  boundary  conditions $Q(\lambda,\lambda)=0$  and
$Q(r,\lambda \to \infty)=1$ are satisfied. The survival probability is
therefore given by
\begin{equation}
\label{3solnr}
Q(r,\lambda )= \frac{1}{r}\sum_{n=-\infty}^\infty \left[ r\cosh (nr)
-2n\lambda \sinh (nr) \right] \rme^{-n^2 \lambda }.
\end{equation}
For a particle starting at the origin we take the limit $r\to 0$, 
giving
\begin{eqnarray}
\label{3soln0}
Q(0,\lambda )& = & \sum_{k=-\infty}^{\infty}\left(1-2\lambda k^2 \right)
\,\rme^{-k^2 \lambda} \\
& \sim & 1 - 2(2\lambda-1)\,\rme^{-\lambda},\ \lambda \to \infty.
\label{3soln1}
\end{eqnarray}
To obtain an expression suitable for extracting the behaviour at 
small $\lambda$, we take the Poisson transform of this sum, giving
\begin{equation}
\label{poisson3d}
Q(0,\lambda )=  \frac{2 \pi^{5/2}}{\lambda^{3/2}}
\sum_{k=-\infty}^\infty k^2 \rme^{
  -\frac{\pi^2 k^2}{\lambda }} \sim \frac{2 \pi^{5/2}}{\lambda^{3/2}} 
\rme^{  -\frac{\pi^2}{\lambda }},
\end{equation}
for $\lambda \to 0$.

For general space dimension $d$, the backward Fokker-Planck method 
does not seem to be useful. The reason for this will become clear in 
the following section.

\section{Survival probability in a ${\bf d}$--dimensional expanding 
sphere}
\label{sec:FFPE}
We approach  the problem in  general dimension by solving  the forward
Fokker--Planck equation for the probability density, $p({\bf r}, t|{\bf
r}_0,  0)$, defined  as  the probability  density  that the  particle,
starting at  position ${\bf r}_0$  within the sphere of  radius $L_0$,
still survives  (has not yet  reached the absorbing boundary)  at time
$t$, and is currently at  position ${\bf r}$. It satisfies the partial
differential  equation $\partial p/\partial  t =  D \nabla^2  p$.  The
boundary condition is $p({\bf r}, t|{\bf r}_0,0) = 0$ for $|{\bf r}| =
L_0 + ct$,  and the initial condition is $p({\bf  r}, 0|{\bf r}_0,0) =
\delta^d({\bf r} - {\bf r}_0)$.

In generalised polar coordinates, the Fokker-Planck equation reads
\begin{equation}
\label{dffpeO}
\frac{\partial p}{\partial t}=D \left( \frac{\partial^2 p}{\partial
 r^2}+ \frac{d-1}{r}\frac{\partial p}{\partial r}+\mathcal{L}_\Omega
 p \right),
\end{equation}
where  $\mathcal{L}_\Omega$   is  a  generalised   angular  derivative
operator.  From  the  spherical   symmetry  of  the  problem,  $p({\bf
r},t|{\bf r}_0,0)$ only  depends on $r \equiv |{\bf  r}|$, $r_0 \equiv
|{\bf r}_0|$,  and the angle between  ${\bf r}$ and  ${\bf r}_0$.  The
problem can be simplified by  choosing the direction of ${\bf r}_0$ as
the principal polar axis, and  integrating out the angular degrees of
freedom by defining
\begin{equation*}
\bar{p}(r,t)=\frac{1}{S_d} \int_\Omega \D \Omega\,
p({\bf r},t|{\bf r}_0,0),
\end{equation*}
where     $d\Omega$    is     an    element     of     solid    angle,
$S_d=2\pi^{d/2}/\Gamma(d/2)$  is the  integral over  the  solid angle,
i.e.\ $S_d$ is the surface area  of the unit sphere in $d$ dimensions,
and the dependence of $\bar{p}(r,t)$ on $r_0$ is implicit.

The differential equation for $\bar{p}$ reads
\begin{equation}
\label{dffpe}
\frac{\partial \bar{p}}{\partial t}=D \left( \frac{\partial^2 \bar{p}}
{\partial r^2}+ \frac{d-1}{r}\frac{\partial \bar{p}}{\partial r} \right).
\end{equation}
The initial condition is 
\begin{equation}
\bar{p}(r,0)= \frac{\delta(r-r_0)}{S_d r_0^{d-1}}\ .
\label{ic} 
\end{equation}

Our method of solution for the moving boundary problem is motivated by
the  solution for  a fixed  absorbing boundary  at  $r=L_0$. Separable
solutions of Eq.~(\ref{dffpe}), regular at the origin, have the form
\begin{equation*}
\bar{p}_k(r,t) = \frac{J_\nu(kr)}{r^\nu}\,\rme^{-k^2 Dt},
\end{equation*}
where  $J_\nu(x)$   is  the  Bessel  function  of   order  $\nu$,  and
$\nu=(d-2)/2$.  The  absorbing boundary condition, $\bar{p}(L_0,t)=0$,
selects  a discrete  set of  $k$-values,  $k_n =  \jroot /L_0$,  where
$\jroot$  is  the $n$th  positive  zero  of  $J_\nu(x)$, to  give  the
discrete set of solutions
\begin{equation}
\bar{p}_n(r,t) = \frac{J_\nu(\jroot r/L_0)}
{r^\nu}\,\rme^{-(\jroot)^2 Dt/L_0^2},
\end{equation}

Our  trial  solution  of   Eq.~(\ref{dffpe})  replaces  $L_0$  by  the
time--dependent    $L(t)=L_0+ct$,   and    $t/L_0^2$    by   $\int_0^t
dt'/L^2(t')$.   We also  multiply the  static solution  by  an unknown
function of $r$ and $t$, to give
\begin{equation}
\label{trial}
\bar{p}(r,t) = g(r,t)\frac{J_\nu \left( \frac{\jroot r}{L(t)} \right) }{r^\nu} 
\exp \left\{ -(\jroot)^2 D \int_0^t \frac{\D t'}{L^2(t')} \right\}.
\end{equation}
At  first sight  it seems  that we  have simply  replaced  one unknown
function, $\bar{p}(r,t)$,  by another,  $g(r,t)$. This is,  of course,
true. However,  the resulting equation for $g(r,t)$ simplifies greatly
for the case of interest, namely $L(t) = L_0+ct$, when it can be exactly
solved. The  same trial solution, Eq.\ (\ref{trial}),  with $n=1$ (the
lowest mode) and  $g(r,t) \to g(t)$, was used  by Krapivsky and Redner
\cite{krap} as a general technique for obtaining an {\em approximate}
solution in the adiabatic limit where the wall is moving slowly.

On substituting Eq.\ (\ref{trial}) into Eq.\ (\ref{dffpe}) we obtain 
the following equation for $g(r,t)$:
\begin{eqnarray}
&& \frac{\partial_{rr}g}{g} + \frac{1}{r}\frac{\partial_r g}{g} 
- \frac{1}{D}\frac{\dot{g}}{g} \nonumber \\
&& +\left(\frac{\jroot}{L}\right)\left(\frac{r\dot{L}}{DL} 
+ 2\frac{\partial_r g}{g}\right)
\frac{J'_\nu(\jroot r/L)}{J_\nu(\jroot r/L)} = 0\ ,
\end{eqnarray}
where dots indicate time  derivatives and $J'_\nu(x) = dJ_\nu/dx$.  We
seek a solution in which  the terms that involve Bessel functions, and
the terms  that do not, separately  vanish. Such a  solution exists if
the two equations
\begin{eqnarray}
\label{gdiff}
\dot{g} & = & D\,\left(\partial_{rr} g + \frac{1}{r}\partial_r g\right) \\
\partial_r g & = & -\left(\frac{\dot{L}r}{2DL}\right)\,g
\label{g2}
\end{eqnarray}
are both satisfied. Eq.\ (\ref{g2}) can be integrated immediately to give
\begin{equation}
g(r,t) = A(t)\,\exp\left(-\frac{\dot{L}r^2}{4DL}\right),
\label{g}
\end{equation}
where $A(t)$ is an arbitrary function. Substituting this result into 
Eq.\ (\ref{gdiff}) gives
\begin{equation}
\dot{A} = \left(\frac{\ddot{L}r^2}{4DL} - \frac{\dot{L}}{L}\right)\,A\,.
\label{A}
\end{equation}
But  $A(t)$ depends  only on  $t$, and  not on  $r$. For  a consistent
solution,  therefore, we  require $\ddot{L}=0$,  i.e. \  the absorbing
boundary must move at constant  speed. Solving Eq.\ (\ref{A}) for this
case gives
\begin{equation}
A(t) = K/L(t)\,
\label{Afinal}
\end{equation}
where $K$ is an arbitrary constant. 

The general solution for the case $L(t)  = L_0 + ct$ is obtained as an
arbitrary  superposition of  separable solutions,  combining equations
(\ref{trial}), (\ref{g}) and (\ref{Afinal}):
\begin{equation}
\label{ddgensoln}
\bar{p}(r,t)=\sum_n \frac{a_n }{L(t) r^\nu} J_\nu \left( \frac{\jroot r}{L(t)}
  \right) \rme^{  - \frac{(\jroot)^2 D t}{L_0 L(t)} -\frac{cr^2}{4DL(t)} },
\end{equation}
the summation being over all positive zeros of $J_\nu(x)$. 

As usual, the amplitudes $a_n$ are determined by the initial condition, 
Eq.\ (\ref{ic}), exploiting the orthogonality property of Bessel
functions:
\begin{equation*}
\int_0^{L_0} r\D r\; J_\nu \left( \frac{\jroot r}{L_0}\right) J_\nu 
\left( \frac{\alpha_\nu^m r}{L_0}\right)  
= \frac{L_0^2}{2} [J_{\nu+1}(\alpha_\nu^m)]^2 \delta_{nm}.
\end{equation*}
The exact solution is thus
\begin{eqnarray}
\label{ddsoln}
\bar{p}(r,t)&=& \frac{2}{L_0L(t)(rr_0)^\nu S_d} 
\sum_n \frac{ J_\nu \left( \frac{\jroot r}{L(t)}\right) 
J_\nu \left( \frac{\jroot r_0}{L_0}\right) }
{[J_{\nu+1}(\jroot)]^2 } \nonumber \\ 
&& \times \; \rme^{-\frac{(\jroot)^2 D t}{L_0 L(t)} 
-\frac{c}{4D}\left(\frac{r^2}{L(t)} - \frac{r_0^2}{L_0}\right)}
\end{eqnarray}

Eq.\  (\ref{ddsoln}) represents  our most  complete result.  Note that
$\bar{p}(r,t)$ depends on  the constants $r_0$, $L_0$, $c$  and $D$ as
well as  $r$ and  $t$. To  make contact with  our earlier  results for
$d=1$   and  $d=3$,   we  now   compute  the   infinite-time  survival
probability,  for  given starting  radius  $r_0$,  by  letting $t  \to
\infty$ and integrating over the final coordinate $r$.

\section{The infinite-time survival probability}
\label{sec:infinite}
To find the probability that the particle survives for infinite time, 
we integrate the probability density over the whole domain. We introduce 
the dimensionless variables, $\rho = cr_0/D$ and $\lambda = cL_0/D$, in  
terms of which the infinite-time survival probability can be expressed:
\begin{eqnarray}
Q(\rho,\lambda) & = & \lim_{t\to \infty}
\frac{2e^{\rho^2/4\lambda}}{L_0L(t)r_0^\nu}
\sum_{n=1}^\infty \frac{J_\nu\left(\frac{\jroot\rho}{\lambda}\right)}
{J_{\nu+1}^2(\jroot)} \nonumber \\ 
&& \times e^{-\frac{(\jroot)^2}{\lambda}}\,I_n(t) 
\label{Q}
\end{eqnarray}
where
\begin{equation}
I_n(t) = \int_0^{L(t)}\frac{r^{d-1}\D r}{r^\nu}
J_\nu\left(\frac{\jroot r}{L(t)}\right)\,e^{\frac{-cr^2}{4DL(t)}}.
\end{equation}
To extract the large-$t$  behaviour, we change variables to $x=r/L(t)$
in  the integral. Then  for large  $t$ the  integral is  dominated by
small values  of $x$ and  the Bessel function  can be replaced  by its
small-argument form $J_\nu(z)  \sim (z/2)^\nu/\Gamma(\nu+1)$, to give,
for $t \to \infty$,
\begin{equation}
I_n(t) \sim 2^{d-1}\left(\frac{D}{c}\right)^{d/2}\,
\left(\frac{\jroot}{2}\right)^\nu\,L(t)\ .
\end{equation}
Putting this result into Eq.\ (\ref{Q}) gives our final result
\begin{equation}
\label{ddQ}
Q(\rho,\lambda)=\frac{2^{\nu+2}}{\rho^\nu
  \lambda}\,\rme^{\rho^2/4\lambda} \sum_{n=1}^\infty
\frac{ (\jroot)^\nu J_\nu \left( \jroot \frac{\rho}{\lambda}\right)}
{[J_{\nu+1}(\jroot)]^2} \rme^{-\frac{(\jroot)^2}{\lambda} },
\end{equation}
where we recall that $\rho=cr_0/D$, $\lambda=cL_0/D$, and $\nu=d/2-1$.

It is interesting to consider the special case of a particle starting 
at the origin. Taking the limit $\rho \to 0$ in Eq.\ (\ref{ddQ}) gives 
\begin{equation}
\label{ddQ0}
Q(0,\lambda)=\frac{4}{\Gamma (\nu+1)
  \lambda^{\nu+1}} \sum_{n=1}^\infty\frac{ (\jroot)^{2\nu} }{
  [J_{\nu+1}(\jroot)]^2} \rme^{-\frac{(\jroot)^2}{\lambda} }.
\end{equation}
The limiting form for small-$\lambda$ is given by the first term in 
the sum. 

We  now  compare  our  results  with  those  obtained  using  backward
Fokker-Planck methods  in section  \ref{sec:BFPE}. The case  $d=1$ and
$d=3$  correspond   to  $\nu=-1/2$  and   $\nu=1/2$  respectively.  The
functions  $J_{-1/2}(z)$  and  $J_{1/2}(z)$  have  positive  zeros  at
$z=(2n-1)\pi/2$       and      $z=n\pi$       respectively,      while
$J_{1/2}^2[(2n-1)\pi/2]=4/[\pi^2(2n-1)]$    and   $J_{3/2}^2[n\pi]   =
2/(\pi^2  n)$. It  is the  fact that  the Bessel  functions  zeros are
uniformly spaced in $d=1$ and  $d=3$ that makes these cases especially
simple.

For $d=1$, the final result, Eq.\ (\ref{ddQ0}), can be written as
\begin{eqnarray}
Q(\rho,\lambda) &=& e^{\rho^2/4\lambda}\sqrt{\frac{\pi}{\lambda}}
\sum_{n=-\infty}^\infty \cos\left((2n-1)\pi\rho/2\lambda\right) 
\nonumber \\
&& \times e^{-(2n-1)^2\pi^2/4\lambda}\ ,
\label{poisson1}
\end{eqnarray}
while for $d=3$ it takes the form
\begin{eqnarray}
Q(\rho,\lambda) &=& e^{\rho^2/4\lambda}\frac{2\pi^{3/2}}{\rho\sqrt{\lambda}}
\sum_{n=-\infty}^\infty n\sin(n\pi\rho/\lambda) \nonumber \\
&& \times e^{-n^2\pi^2/\lambda}\ .
\label{poisson3}
\end{eqnarray}
For   the   special   case   $\rho=0$,   these   results   reduce   to
Eqs.(\ref{poisson})   and  (\ref{poisson3d}).   One   can  show   this
correspondence holds for general $\rho$ by using the Poisson summation
formula (with  $\rho \to  y$ in $d=1$  and $\rho  \to r$ in  $d=3$) to
transform   Eq.\   (\ref{ddQ})   into   Eqs.\   (\ref{coshsoln})   and
(\ref{3solnr}) for $d=1$ and $d=3$ respectively.

It  is  not  simple to  use  the  Poisson  summation formula  on  Eq.\
(\ref{ddQ}) for general $d$, since we do not have explicit expressions
for $\jroot$  except for $\nu  = \pm 1/2$.  This means that it  is not
straightforward to analytically extract the large $\lambda$  limit  of
$Q(0,\lambda)$. In the final part  of this paper, therefore, we extend
the  ``fast approximation'' of  ref.~\cite{krap} to  general dimension
$d$ in order to investigate the large-$\lambda$ behaviour.

\section{Approximate solution for fast expansion (large $\lambda$)}
\label{sec:fast}
For  a  rapidly  expanding  cage  in  $d$  dimensions  we  follow  the
approximate  method used  in \cite{krap}.   This approach  consists of
equating the  loss of survival probability to  the outward probability
flux at the boundary. For a particle starting at the origin, we define
$Q(t)$ to  be its survival  probability at time  $t$.  If the  cage is
expanding rapidly, the probability  flux through the boundary is small
and we can use the free form for the diffusion propagator:
\begin{equation*}
p(r,t)=Q(t)\,\frac{\exp \left(-r^2/4Dt\right)}{(4\pi Dt)^{d/2}},
\end{equation*}
where $r$ is the distance from the origin at time $t$. We equate the 
rate of decrease of $Q$ to the flux through the boundary: 
\begin{equation*}
\frac{\D Q}{\D t}= S_d L(t)^{d-1} D 
\left.\partial_r p(r,t) \right|_{r=L(t)},
\end{equation*}
or
\begin{eqnarray}
-\ln Q(t) &=& \frac{1}{\Gamma (d/2) (4D)^{d/2}} \int_0^t \D u\, 
 \frac{(L_0+cu)^d}{u^{\frac{d}{2}+1}} \nonumber \\ 
&& \times \exp \left(-\frac{(L_0+cu)^2}{4Du} \right).
\end{eqnarray}
Setting $t=\infty$, we can evaluate the integral for large $\lambda$ 
using the method of steepest descents to obtain the infinite-time result
\begin{eqnarray}
Q(\infty) &=& \exp \left(- \frac{2\sqrt{\pi}\lambda^{\frac{d-1}{2}}}
{\Gamma(d/2)} \; \; \rme^{-\lambda} \right), \nonumber \\
& \sim & 1 - \frac{2\sqrt{\pi}\lambda^{\frac{d-1}{2}}}
{\Gamma(d/2)} \; \; \rme^{-\lambda}\ , \ \ \ \lambda \to \infty,
\label{largeapprox}
\end{eqnarray}
which, for the cases $d=1$ and $d=3$, agrees with the large-$\lambda$ 
limits given in  Eqs~(\ref{qlambda}) and (\ref{3soln1}) respectively.

\section{Summary}
\label{sec:summary}
In  this paper  we have  derived exact  results,  in arbitrary  space
dimension $d$,  for the survival  probability of a  particle diffusing
inside a uniformly  expanding cage that acts an  an absorbing boundary
for  the  particle.  We  found  very simple  forms  for  the  survival
probability  in   one  and  three  dimensions  by   using  a  backward
Fokker-Planck approach.  Our solution for general  $d$, however, using
forward Fokker-Planck  methods, indicates  that the simplicity  of the
solutions  in one  and three  dimensions is  fortuitous. Our method of
solution seems to  be restricted to the case  where the boundary moves
at  constant  speed, though  it  will  be  interesting to  pursue  the
question of whether there are any other soluble cases.
  
\bigskip
\begin{small}
\noindent{\bf ACKNOWLEDGEMENTS}
\end{small}
\medskip

The work of RS was supported by EPSRC.

\end{document}